# Using GSM SMS Controller Alarm Configurator to Develop Cost Effective Intelligent Fire Safety System in a Developing Country

Ebenezer Narh Odonkor[1], Willie K. Ofosu[2] and Kingsley Nunoo[1]

[1] Faculty of Engineering, Department of Electrical and Electronic Engineering,
Takoradi Technical University, Takoradi – Ghana.
Email: [1]ebenezer.narh.odonkor@tpoly.edu.gh.,[1]snsnunoo@gmail.com.
[2] Penn State Wilkes-Barre, 44 University Drive, Dallas, PA 18612.
Email:[2]wko1@psu.edu

*ABSTRACT*

*Electricity supply to facilities is essential, but can cause fires when care is not taken, and can destroy the properties within some few minutes. The need for fire protection is therefore essential. This paper addresses this pertinent issue facing Ghana as a developing country. At the fire outbreak, the designed system responds to the smoke and cuts off the electricity supply. When a fire is detected, an alarm is turned on and Short Messaging Service (SMS) alert is sent to the owner. After ten to fifteen seconds, the system, resends SMS to the owner and the National Fire Service (NFS) safety officer giving the exact location. The call alert is sent to the owner of a facility and fire safety personnel respectively, when the system is not reset. A microcontroller serves as the command center of the cost effective, intelligent fire safety system. The circuit is designed and simulated using Proteus software, and programing was done using Global System for Mobile Communications (GSM) SMS Controller Alarm Configurator software. The prototype was constructed and tested in real-time. The proposed system is cost effective and will help policy makers, and save lives and property when implemented.*

*KEYWORDS*

*Microcontroller, GSM, SMS, Call alert, and Smoke Sensor.*

## 1. Introduction

Some of the most commonly known causes of fire outbreak accidents in developing countries like Ghana include faulty electrical cables, gas leakage and human error [1]. Occupants' negligence of fire safety rules also contributes to fires getting started. According to [2] occurrence of a fire in a domestic environment, companies (factories), commercial buildings, government facilities, and important properties have economic consequences that affect nations, regions, communities, and governments all over the world, most especially developing countries. This point is echoed by [3]. There is, therefore, the need for early fire detection, and providing automatic protection remotely by switching off the electrical supply to the facility. Switching off the power supply from a facility is one of the ways in preventing spread of fire outbreak hence the need for this project.

This paper proposes an early fire detection system using a smoke detector to shut down the power supply to a facility when smoke above the threshold of the smoke sensor is detected. A notification of fire detection, and electrical supply shutdown to property is communicated to the owner and fire safety personnel automatically through Global System for Mobile Communications (GSM) for prompt actions to be taken.

## 2. BACKGROUND OF THE STUDY

Home automation and network system requirements have increased with the demand for smarter homes around the world [4][5]. Smart home appliances are therefore connected to the home network to monitor and control the appliances from inside the house, factory, commercial and government buildings remotely using GSM technology. Almost all homes, offices, factories (industries), government buildings in both developed and developing countries around the world are supplied with high voltage alternating current (HVAC) power source. In recent years, some of these buildings are connected using automation systems to detect anti-crime and fire-detection to improve convenience and safety for occupants [4] [5]. Home electrical equipment such as main switches, distribution boards, circuit breakers, light, fans, televisions, socket outlets, heaters, air conditioners, refrigerators and fire detectors, are directly connected to the controllers of HAVC or fire-detection systems for remote notification. This new technology is only installed in a few facilities in Ghana [3]. Information obtained indicates that 5,531 fire outbreak cases were recorded by the Ghana National Fire Service (GNFS) across the country in 2018. From this, 43 people died and a total of 57 persons got injured, and the fires caused damage to property worth GH¢36 million (about $7.34 million). In 2017, a total of 4,544 cases of fire outbreaks in various facilities was recorded. Thousands of lives and properties are lost due to fire outbreaks in Ghana [3]. This necessitates the need for developing a cost effective, intelligent fire safety system in Ghana using GSM SMS controller alarm Configurator. This proposed system can cut off electricity supply to facilities automatically any time smoke above a certain threshold is detected. The system also has the ability to call the owner(s) and national fire service personnel. Some data gathered based on the trend of fire outbreak in Ghana are presented in tables and graphical forms as shown below.

Table 1 below shows the frequency of occurrences of outbreak of fire in the various regions in Ghana in the year 2018

Table 1: Occurrence of fire outbreak in various regions in Ghana.

| Number of Regions | Regions | Occurrences of Fire Outbreak |
|---|---|---|
| 1 | Greater Accra Region | 522 |
| 2 | Ashanti Region | 542 |
| 3 | Central Region | 294 |
| 4 | Eastern Region | 287 |
| 5 | Brong Ahafo Region | 284 |
| 6 | Western Region | 243 |
| 7 | Northern Region | 201 |
| 8 | Upper East Region | 145 |
| 9 | Volta Region | 146 |
| 10 | Upper West Region | 64 |

Source: [2]

International Journal of Computer Science, Engineering and Applications (IJCSEA) Vol.10, No.2/3, June 2020

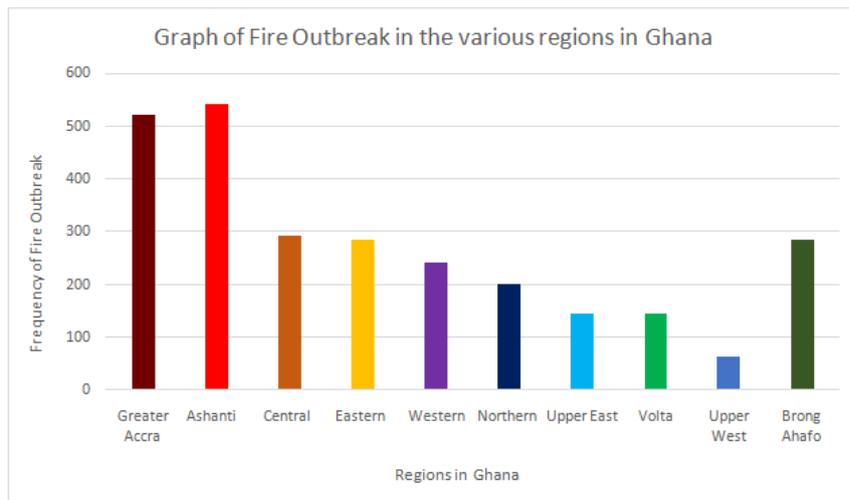

Figure 1: Graph of Fire Outbreak in the various regions in Ghana

Table 2: Fire outbreak in various sectors in Ghana

| Number of Regions | Type of Fire | Frequency of Fire Outbreak |
|---|---|---|
| 1 | Domestic | 1,794 |
| 2 | Industrial | 110 |
| 3 | Vehicular | 540 |
| 4 | Institutional | 120 |
| 5 | Electrical | 544 |
| 6 | Bush fires | 859 |
| 7 | Unclassified | 313 |

Source: [2]

Figure 2 shows the graphical notation of figures in Table 2 which gives the frequency of fire outbreak in various sectors in Ghana.

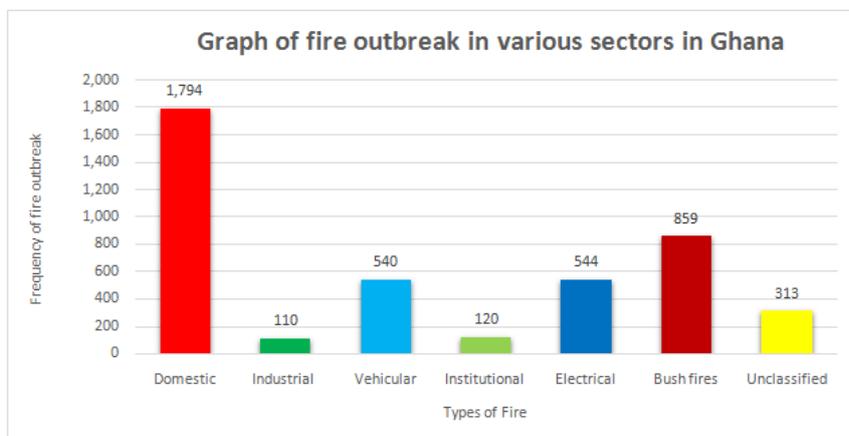

Figure 2: Graph of fire outbreak in various sectors in Ghana

Based on the information in Table 1 above, 542 fire outbreaks were recorded in Ashanti, in Greater Accra 522 fire cases were recorded. In Upper West Region, 64 fire outbreaks were recorded. This shows that Upper West region recorded the lowest fire outbreaks in Ghana as compared to other regions during the year under review. From the graph in Figure 2 above, Domestic premises recorded 1,794 outbreaks indicating that Domestic premises in Ghana are the highest areas prone to fire cases, followed by bush fire where 859 cases were recorded. Electrical fire recorded were 544, vehicular fire occurrences were 540 with industrial facilities, recording 110 cases of fire outbreak. The statistics from the graph shows clearly that industrial facilities represent the lowest rate of fire occurrence in Ghana.

## 3. RELATED WORKS

[6] Developed a Fire Alarm and Detection System based on text message technology with an embedded GSM module. Their proposed system is able to detect fire, give an alarm alert and send SMS to homes, and office owners. Their results show that implementing this will reduce uncontrolled fires by 50% due to the warning of dangerous conditions. These authors' system is deficient in shutting down electrical power supply to a facility.

[7] Designed a smoke detector alarm system using GSM communication system. However, their proposed system cannot turn off the electrical power supply to a building automatically when a fire outbreak occurs.

[8] Developed a low cost, smart home and industrial automated security systems based on GSM technology. In their system architecture development, Arduino UNO, Temperature sensor (Resistance Temperature Detector, Thermistors and Thermocouples) were used. Even though their system can send SMS when fire is detected, it does not cut off the entire power supply to the facility to minimize the damage.

Intelligent gateway interfacing with fire monitoring workstation using Noti-Fire-Net (NFN) Gateway was proposed by [9] which also support full panel programming using embedded C (program) and network diagnostics for home automation.

From the above research works, it is clear that research has been carried out on smart home automation for fire detection using GSM technology. It is also clear that some research has been carried out on developing a cost effective, intelligent fire safety system in developing countries to control electrical power supply to the facility in the occurrence of fire outbreak. This paper presents an intelligent fire safety system that can automatically cut off facility supply when a fire outbreak is detected and send SMS call alert to owners and fire safety personnel during domestic, commercial and industrial buildings fire outbreaks in Ghana.

## 4. BLOCK DIAGRAM

Figure 3 below shows the block diagram of the proposed GSM SMS Controller Alarm Configurator with Intelligent Fire Safety System for buildings supplied with alternating current (AC) intended for developing countries. The block diagram of the proposed system comprises of a microcontroller, GSM module, power supply, and smoke detector. Also shown in the block diagram is alarm unit, emergency light, DC backup supply, power supply tripper, mobile phone, loads, contactor, and distribution board. The programing done by controlling the proposed system is stored in the brain of the circuit (microcontroller ATMEGA328)

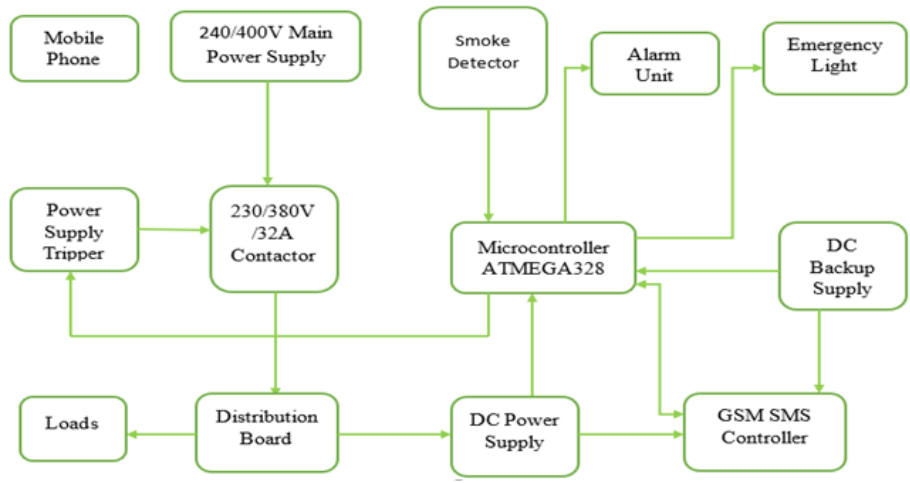

Figure 3: Block Diagram of the proposed GSM SMS Controller Alarm Configurator with Intelligent Fire Safety System

## 5. DESCRIPTION OF BLOCK DIAGRAM

Power is supplied to the facility (building) through a contactor to the consumer unit before connecting it to the final circuits for distribution to the entire load in the building under normal conditions. The smoke detector sends a signal to the integrated circuit (IC) under normal and abnormal conditions. The GSM SMS controller unit is powered by 12VDC and also has backup power supply inbuilt. An emergency light is connected to the IC circuitry to turn on an emergency light for occupants to evacuate buildings at night when a fire outbreak occurs. When the smoke detector detects smoke above the threshold, a signal is sent to the microcontroller unit from the smoke detector. A power shutdown command is sent from the IC to the power supply tripper to cut off the power supply from the facility. The microcontroller again sends a signal to the GSM module and the fire alarm circuitry, and a call alert is sent to the owner of the facility, and the fire service personnel automatically.

## 6. CIRCUIT DESIGN

Figure 4 below shows the circuit diagram of the proposed system using GSM SMS controller alarm Configurator in developing the system.

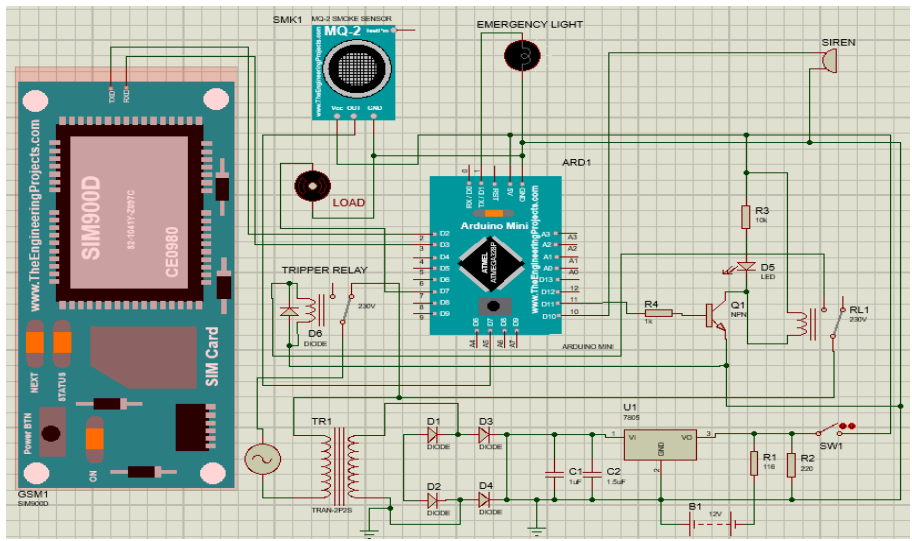

Figure 4:The circuit diagram of the proposed GSM SMS controller alarm Configurator with intelligent fire safety system.

The heart of the circuit is the S150 alarm controller, which coordinates all the information from and to all the various peripherals that have been connected to it. The microcontroller module is interfaced to the hardware universal serial bus port of the controller through a level converter designed with two (2) output relay 240VAC3A switch contacts for high voltage switching. The controller has eight digital input signal commands. It has a dual 12Vdc output power source which is used to power the fire alarm siren when the smoke detector detects smoke more than the threshold settings of the detector. The controller takes its AC power source from the main power supply when the switch is closed. The positive terminal A1 of the output relay are connected to the positive terminal of the main supply breaker and the negative terminal A2 is connected to the neutral point of the supply. The tripper is coupled to the breaker (Distribution board) which controls the loads. Appliances in the building are supplied with power when the relay is energized through the switch wire connected to contact one (1) of the controller.

The smoke detector is connected to terminal D0 of the microcontroller (IC). The intelligent fire safety system controller alarm Configurator is configured with a programmable software tool. The IC is programmed to detect smoke, activate an alarm, activate supply cutoff command, send an SMS and activate a call alert. When smoke is detected, the supply tripper is activated, resulting in the total power cutoff. Since this occurrence could result in total darkness, especially at night, an emergency light is provided to give illumination for the occupant evacuation. The siren (fire alarm) is triggered at the activation of the system, thereby sounding an alarm repeatedly until the smoke detector is deactivated or reset. During the active period, SMS is sent to the designated numbers sequentially, and if the system is not reset within 15 seconds, the system automatically calls the owner of the facility and the Ghana National Fire Service office respectively for prompt action to be taken.

## 7. RESULT AND DISCUSSION

Figure 5 below shows the final packaging of the prototype, and Figure 6 shows SMS received from the artifact when the power was cut off and the fire alarm was activated.

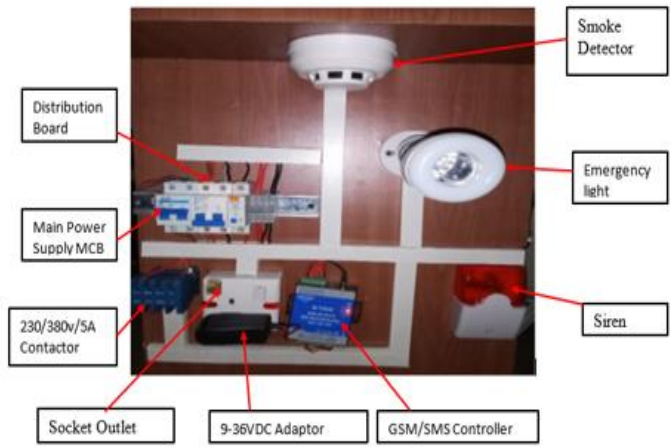 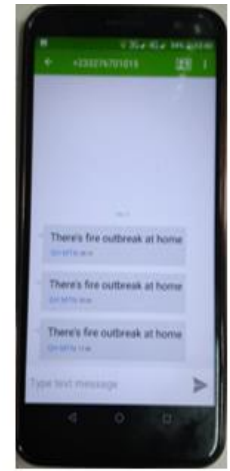

Figure 5: Package of the artifact    Figure 6: Test message

In the situations presented in Section II, Background of the Study above, lives and properties were lost, causing millions of Ghana cedis. In view of this great loss caused by fire, a prototype of cost effective, an intelligent fire safety system using GSM SMS controller alarm Configurator for developing countries is proposed. After the construction of the prototype, it was tested in a real-time basis where the power supply to the artifact automatically cut off, an alarm was activated, emergency light was turned on, an SMS and call alert was sent to the programed numbers in sequence (in this case the owner of the facility and Ghana National Fire Service office respectively) to give notification of a fire outbreak for prompt action to save lives and property.

## 8. CONCLUSION

Safety at home is a key in developing a society, community and a nation at large, hence the increase of the use of home automation technologies in monitoring home facilities. The increasing occurrence of fire outbreaks in facilities in Ghana is alarming. An intelligent fire safety system has been developed for buildings' power supply control using a smoke detector with GSM SMS controller alarm technology system. The system was designed using Proteus Software. Microcontroller was used as the brain of the prototype to store the programming code which is used to control the prototype. Programming was done using S150 GSM SMS alarm Controller software. The construction of the proposed cost effective, intelligent fire safety system is simple and has more advantages. It can be implemented in institutions, market places and industrial facilities, especially in developing countries like Ghana. The system sends an alert to a facility owner(s) and fire safety officers (GNFS) to provide preventive measures to avert danger of fire outbreak at domestic, commercial, institutional, offices, warehouses and industrial facilities. It is therefore recommended that this system may be implemented in all facilities in developing countries around the world where fire hazards of this nature can be a problem. Implementing this technology in building structures in Ghana or anywhere in the world will help to reduce fire outbreaks. It will save lives, and the cost to facility owners and governments as a whole.